\documentclass[conference]{IEEEtran}

\DeclareFontShape{OT1}{cmtt}{bx}{n}{<5><6><7><8><9><10><10.95><12><14.4><17.28><20.74><24.88>cmttb10}{}

\newcommand{\eifkw}[1]{{\bf #1}}
\newcommand{\eif}[1]{\textsf{#1}}

\newcommand{\eifcomment}[1]{{-}{-}\textsf{#1}}

\title{Usability of AutoProof: \\ a case study of software verification}

\author{\IEEEauthorblockN{Mansur Khazeev\IEEEauthorrefmark{1},
Victor Rivera\IEEEauthorrefmark{2}, 
Manuel Mazzara\IEEEauthorrefmark{3} and
Alexander Tchitchigin\IEEEauthorrefmark{4}}
\IEEEauthorblockA{Innopolis University,
Software Engineering Lab.\\
Innopolis, Russia\\
Email: \IEEEauthorrefmark{1}m.khazeev@innopolis.ru,
\IEEEauthorrefmark{2}v.rivera@innopolis.ru,
\IEEEauthorrefmark{3}m.mazzara@innopolis.ru,
\IEEEauthorrefmark{4}a.chichigin@innopolis.ru}}

\usepackage{graphicx}
\usepackage{amssymb}
\usepackage{lineno}
\usepackage{verbatim}
\usepackage{listings}
\usepackage{todonotes}
\usepackage{cite}
\usepackage{zed-csp}

\begin{document}

\maketitle
\thispagestyle{empty}
\pagestyle{empty}

\begin{abstract}

Many verification tools come out of academic projects, whose natural constraints do not typically lead to a strong focus on usability. For widespread use, however, usability is essential. Using a well-known benchmark, the Tokeneer problem, we evaluate the usability of a recent and promising verification tool: AutoProof. The results show the efficacy of the tool in verifying a real piece of software and automatically discharging nearly two thirds 
of verification conditions. At the same time, the case study shows the demand for improved documentation and emphasizes the need for improvement in the tool itself and in the Eiffel IDE.

\end{abstract}

\section{Introduction}
\label{sec:intro}


Modern systems have grown fast in complexity, and demand for quality. Focus on quality, in turn, demands stronger attention to the entire development life-cycle. With the tendency to reuse and integration, the need for software quality is even more important since applications and components have to rely on each other with partial knowledge of the implementation and based on interface only. 


Tools for software verification allow the application of theoretical principles in practice, in order to ensure that nothing bad will ever happen (safety). The extra effort required by the use of these tools is certainly not for free and comes with increased development costs \cite{Meyer:Touch:2009}. There is a common belief in industry that developing software with high level of assurance is too expensive, therefore not acceptable, especially for non safety-critical or financially-critical applications. 

Tools and techniques for the formal development of software have played a key role on demystifying this belief. There are several approaches, for instances abstract interpretation and model checking \cite{Cousot:abstractint:77, modelchecking:clark:2000}, that seek the automation to formally proving certain conditions of systems. However, these techniques tend to verify simple properties only. On the other end of the spectrum, there are interactive techniques for verification such theorem provers \cite{Loveland:theoremprover1978}. These techniques aim at more complex properties, but demand the interaction of users to help the verification. 

Nowadays, there are new approaches that aim at finding a good trade-off between both techniques, e.g. auto-active: users are not needed during the verification process (it is automatically performed); they are required instead to provide guidance to the proof using annotations.  AutoProof \cite{autoproof:julian:15}, is a static auto-active verifier for functional properties of object-oriented programs. Using AutoProof, users write code and equip classes with contracts and annotations to help the tool to prove certain properties. 

The main goal resented in this paper is to provide insights on how easy/difficult is for users (mainly engineers without deep knowledge of formal verification) to use current methodologies and tools for the development of software with high level of assurance, in particular on the use of the AutoProof tool. 

Generally, to prove the correctness of a program one needs some mechanisms to express what the program is supposed to do and clearly state it in the specifications that are used later to verify the program. Eiffel programming language natively supports these mechanisms by means of contracts. Eiffel is an object-oriented programming language which directly implements the concepts of Design-by-Contract (DbC) \cite{Meyer:OOSC:97, Meyer:Touch:2009}. The key concept is viewing the relationship between a class and its clients as a formal agreement, expressing each party's rights and obligations. This is realized equipping methods with pre- and post-conditions, and classes with invariants. The key feature of the Eiffel language is indeed the idea that all the methods might and should contain contracts. 

Contracts and annotations used in Eiffel are used by AutoProof to statically verify the consistency of the classes. To demonstrate the usability of the tool, the Tokeneer project \cite{tok_sources} was implemented in Eiffel and AutoProof was used to verify the consistency of the code. The Tokeneer project is a system specified and implemented by National Security Agency (NSA). Initially, NSA carried out this challenge to prove that it is possible to develop secure systems rigorously in a cost effective manner. Since its development, it became a testing range for different software development methodologies and verification tools. Results of the project are publicly available. This paper reports on the use of AutoProof to verify an Eiffel implementation of Tokeneer and also reports on how easy/difficult is for users to use the tool, e.g. the burden of helping the tool by means of annotations in the code.

The rest of the paper is organized as follows: Section \ref{sec:pre} introduces the Tokeneer project, Eiffel and the AutoProof tool. Section \ref{sec:verification} describes the methodology used to verify the implementation of the Tokeener project. Section \ref{sec:empirical} presents empirical results helping to draw conclusions. Section \ref{sec:related} is devoted to related work and Section \ref{sec:conclusion} concludes and mentions future work.

\section{Preliminaries}
\label{sec:pre}
\subsection{The Tokeneer Project}
In 2002, with the aim to prove/disprove the common believe in industry that development of software of high level of assurance is too expensive and therefore not feasible, the National Security Agency (NSA) asked Altran to undertake a research project to develop part of an existing secure system, the Tokeneer System, in accordance with Altran's Correctness by Construction development process. The system was specified using Z notation \cite{Z-Abrial:80} and implemented in Ada\cite{ada:03}. The project was successfully delivered in 2003 within 260 days of effort, and later, in 2008, all the results were made available by NSA to the software development and security communities in order to demonstrate the possibility to develop secure systems in a cost effective manner. It includes the ``Core'' Tokeneer ID System Software, test cases derived from the system test specification, ``Support'' Tokeneer ID System Software and test tokens and biometric data, project documents. Since the delivery, the Tokeneer project has become a milestone point and a testing range for different verification tools before applying them in industrial projects. Despite the fact that after delivery 4 bugs\footnote{According to \cite{tok_sources}} were found, the system is still deemed to be very secure. 
 
\begin{figure}
\centering\includegraphics[width=0.9\linewidth]{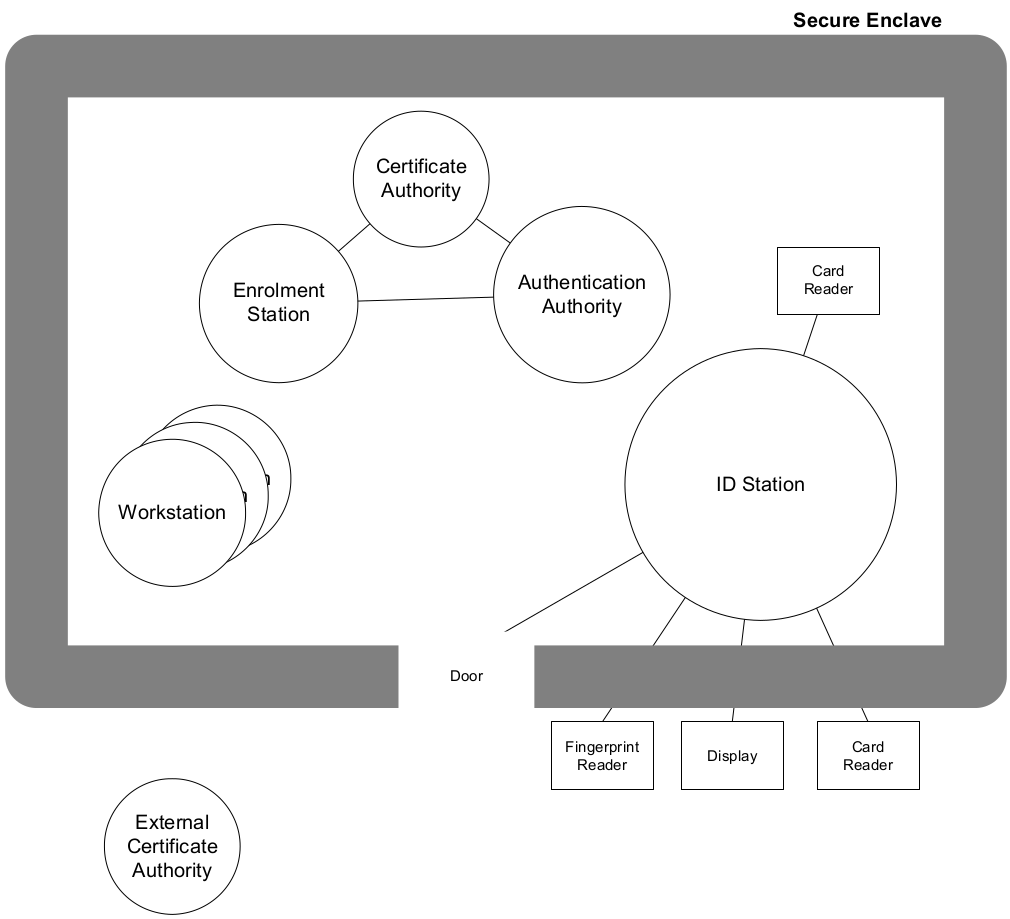}
\caption{The Tokeneer System}
\label{fig:tokeneer}
\end{figure}

Tokeneer is a secure enclave consisting of a set of system components, some housed inside the enclave and some outside, as depicted in Figure \ref{fig:tokeneer}. The ID Station (TIS) is part of the larger Tokeneer system. It has four connected peripherals, namely, a fingerprint reader, a smartcard reader (users use Tokens -smartcards- as identification), a door and visual display. The objective of the enclave is to ensure that anyone who enters the enclave has a proper access, and no one else can access to the enclave. 

In order to ensure the entrance of users to the enclave, TIS implements a series of protocols and checks (the use of smart cards and biometrics) to grant or deny the entrance to it. This paper discusses one of these protocols: the Enrollment to the ID Station. The protocol starts in a state where the user is not enrolled. Users can request enrollment and then insert a FLOPPY (it retains an internal view of the last data written) for the system to proceed. The system reads the information in the floppy and either fails the enrollment process, in which case takes the process to the initial state, or correctly validates the data in the floppy. 


\subsection{Eiffel}

Eiffel is a real complex object oriented programming language that natively supports Design-by-Contract methodology. Users can specify the behavior of Eiffel classes by equipping them with contracts: pre- and post-conditions and class invariants, that are represented as assertions. 

\begin{figure}[!ht]
{\small
\[	
\begin{array}{c@{\hspace*{15pt}}c}
\begin{array}{l}
\eifkw{class}\\
\hspace{0.1cm}\eif{ACCOUNT}\\
\eifkw{create} \eif{ make}\\
\\
\eifkw{feature}~ \eifcomment{ Initialization}\\
\hspace{0.1cm}\eif{make } \eifcomment{Initialize empty account.}\\
\hspace{.2cm}\eifkw{do}\\
\hspace{.3cm}\eif{balance := 0}\\
\hspace{.2cm}\eifkw{ensure}\\
\hspace{.3cm}\eif{balance\_set: balance = 0}\\
\hspace{.2cm}\eifkw{end}\\
\\
\eifkw{feature}~ \eifcomment{ Access}\\
\hspace{0.1cm}\eif{balance}: \eifkw{INTEGER}
\hspace{.3cm}\eifcomment{ Balance of account.}\\
\\
\eifkw{feature}~ \eifcomment{ Element change}\\
\hspace{0.1cm}\eif{deposit }(\eif{amount}: \eifkw{INTEGER})\\
\hspace{.3cm}\eifcomment{ Deposit `amount' on account.}\\
\hspace{.2cm}\eifkw{require}\\
\hspace{.3cm}\eif{amount\_not\_negative}: \eif{amount $>=$ 0}\\
\hspace{.2cm}\eifkw{do}\\
\hspace{.3cm}\eif{balance := balance + amount}\\
\hspace{.2cm}\eifkw{ensure}\\
\hspace{.3cm}\eif{balance\_increased}: \eif{ balance = } \\
\hspace{1.3cm}\eifkw{old} \eif{ balance + amount}\\
\hspace{.2cm}\eifkw{end}\\
\\
\hspace{0.1cm}\eif{withdraw }(\eif{amount}: \eifkw{INTEGER})\\
\hspace{.3cm}\eifcomment{ Withdraw `amount' from account.}\\
\hspace{.2cm}\eifkw{require}\\
\hspace{.3cm}\eif{enough\_balance}: \eif{ amount $<=$ balance}\\
\hspace{.2cm}\eifkw{do}\\
\hspace{.3cm}\eif{balance := balance - amount}\\
\hspace{.2cm}\eifkw{ensure}\\
\hspace{.3cm}\eif{balance\_decreased}: \eif{ balance = }
\eifkw{old} \eif{ balance - amount}\\
\hspace{.2cm}\eifkw{end}\\
\\
\eifkw{invariant}\\
\hspace{0.1cm}\eif{non\_negative\_balance}: \eif{ balance $>=$ 0}\\
\eifkw{end}
\end{array}
\end{array}
\]
} 
\caption{\eifkw{ACCOUNT} Eiffel class}
\label{fig:class}
\end{figure} 

Figure \ref{fig:class} depicts a reduced implementation of a Bank Account. In Eiffel, creation procedures are listed under the keyword \eifkw{create}, for class \eifkw{ACCOUNT}, routine \eif{make} is used as a creation procedure. The class defines a class attribute \eif{balance} to represent the current balance of the account. It also defines two routines (methods), \eif{deposit} and \eif{withdraw}. \eif{deposit} implements a deposit of amount \eif{amount} of money to the account and \eif{withdraw} implements withdrawing money. Eiffel  encourages software developers to express formal properties of classes by writing assertions. Routine pre-conditions express the requirements that clients must satisfy whenever they call a routine. They are introduced in Eiffel by the keyword \eifkw{require}. Routine \eif{deposit} imposes a pre-condition on the call, the client must pass as an argument a non-negative number (i.e. \eif{amount\_not\_negative: amount $>=$ 0}) for the routine to work correctly: a negative value might invalidate the invariant of the class. Routine post-conditions, introduced in Eiffel by the keyword \eifkw{ensure}, express conditions that the routine (the supplier) guarantees on method exit, assuming the pre-condition. Routine \eif{deposit} guarantees that the balance of the account will be the previous value of the balance (expressed in Eiffel by the keyword \eifkw{old}: the value on entrance of the routine) plus the amount being deposited. Routine \eif{withdraw} imposes the constraint to the caller that the argument must be less than or equal to the current balance of the account to avoid having negative value in the balance. The routine ensures that, after execution, the new value of \eif{balance} will be the value on routine entry minus the amount withdrawn.

A class invariant must be satisfied by every instance of the class whenever the instance is externally accessible: after creation, and after any call to an exported routine of the class (public routines). The invariant appears in a clause introduced by the keyword \eifkw{invariant}. Class \eifkw{ACCOUNT}'s invariant imposes the restriction that class attribute \eif{balance} can never be negative (i.e. \eif{non\_negative\_balance: balance $>=$ 0}).

\subsection{AutoProof}
AutoProof \cite{autoproof:julian:15} is a static verifier of contracts for Eiffel programs. It follows the auto-active paradigm where verification is done completely automated, similar to model checking \cite{modelchecking:clark:2000}, but users are expected to feed the classes providing additional information in the form of annotations to help the proof. AutoProof identifies software issues without the need of executing the code, therefore opening a new frontier for ``static debugging'', software verification and reliability, and in general for software quality.

AutoProof verifies the functional correctness of Eiffel classes. It translates Eiffel code to Boogie programs \cite{Rustan:Boogie:08} and calls the Boogie tool to generate verification conditions: logic formulas whose validity entails correctness of the input programs. Finally, retrieves the answer back to Eiffel. AutoProof verifies that routines satisfy pre- and post-conditions, maintenance of class invariants, loops and recursive calls termination, integer overflow and non \eifkw{Void} ($null$ in other programming languages) references calls. The tool also supports most of the Eiffel language constructs: in-lined assertions such as \eifkw{check} ($assert$ in other programming languages), types, multi-inheritance, polymorphism.

\section{Verification of Tokeneer using AutoProof}
\label{sec:verification}
The Tokeneer project was implemented in Eiffel following the specifications file 41\_2.pdf (see \cite{tok_sources}) of the Tokeneer System  and equipping classes with contracts. This research work encompasses only the enrolment process of the whole Tokeneer System therefore it implements only the entities involved in this process.

One of the main parts of TIS is the \eifkw{ID\_STATION} (see Figure \ref{fig:idstation}) - it describes how all  components of the system are related to each other: one of the components is implemented in class \eifkw{INTERNAL\_S} (not shown here) whose responsibility is to keep knowledge of the status of user entry and the enclave and to hold a timeout when relevant; another component is implemented on class \eifkw{FLOPPY} (not shown here) that retains an internal view of the last data written to the floppy as well as the current data on the floppy. \eifkw{ID\_STATION} displays the configuration data on the screen which is implemented in \eifkw{SCREEN\_DISPLAY}. There are a number of messages that may appear on the TIS screen. The Real World types (described in \cite{tok_sources} Specification document, section 2.7.1) of the system such as messages that appear on the display and screen, were implemented all together in class \eifkw{CONST} which implements the constants used in the TIS. And finally, a number of interactions between all these entities within the enclave are implemented in \eifkw{ENCLAVE\_OPERS}.

AutoProof does not make any assumptions out of box therefore users are expected to feed the Eiffel classes for a succeed verification. 
\begin{figure}
{\small
\[
\begin{array}{c@{\hspace*{15pt}}c}
\begin{array}{l}
\eifkw{class}\\
\hspace{0.3cm}\eifkw{ID\_STATION}\\
\hspace{0.5cm}\eif{\ldots Some lines were omitted\ldots}\\
\eifkw{create}\\
\hspace{0.3cm}\eif{make}\\
\eifkw{feature}~ \eifcomment{Initialization}\\
\hspace{0.3cm}\eif{make}\\
\hspace{0.5cm}\eifkw{note}\\
\hspace{0.7cm}\eifkw{status: creator}\\
\hspace{0.5cm}\eifkw{do}\\
\hspace{0.7cm}\eif{\ldots Some lines were omitted\ldots}\\
\hspace{0.5cm}\eifkw{end}\\
\eifkw{end}
\end{array}
\end{array}
\]
} 
\caption{Initialisation of \eifkw{ID\_STATION} Eiffel class.}
\label{fig:initids}
\end{figure} 
This is expressed by means of Eiffel's \eifkw{note} clause. \eifkw{note} clause enables users to attach addition information to the class that is ignored by the Eiffel's compiler. AutoProof uses this information to succeed in the verification. For instance, Autoproof's annotation \eifkw{status} defines which procedure is used to initialize newly created objects: Figure \ref{fig:initids} depicts procedure \eif{make} with annotation \eifkw{note} (e.g. \eifkw{note} \eif{status: creator}) to help Autoproof to discharge the corresponding proof obligations related to creation procedures: the procedure will be called only when an object of this class is being created, AutoProof needs to verify a creation routine only once.

\eifkw{note} clause is also used to define models queries to express the abstract state space of a classes. Model queries are part of model-based contracts to help users to write abstract and concise specifications \cite{mml:polikarpova:10}, they are used to specify the behavior of the class. In Eiffel, this is specified by adding a \eifkw{note} clause at the beginning of the class followed with a keyword \eifkw{model:} and listing one or more attributes of the class. Model queries are also used to describe frame conditions: which allocations are allowed to be modified by procedures.  
 In Eiffel, frame conditions are listed using the \eifkw{modify} clause, which lists the model queries that the feature is allowed to modify, as shown in Figure \ref{fig:modify} (i.e. \eif{modify\_model("current\_display", Current)}).

\begin{figure}
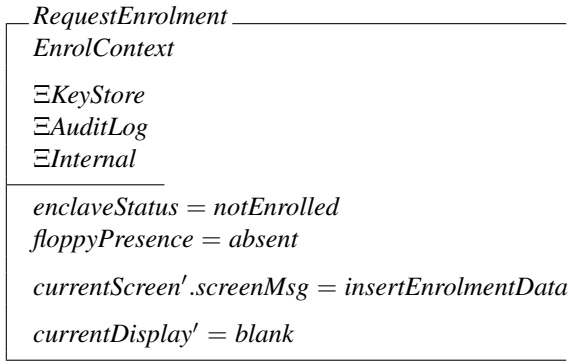

\begin{schema}{RequestEnrolment}
        EnrolContext
\also
        \Xi KeyStore
\\      \Xi AuditLog
\\      \Xi Internal
\where
        enclaveStatus = notEnrolled
\\      floppyPresence = absent
\also
        currentScreen'.screenMsg = insertEnrolmentData
\also
        currentDisplay' = blank
\end{schema}
\caption{Z schema of RequestEnrolment}
\label{fig:schema}
\end{figure}
According to RequestEnrolment (a Z-schema that is a part of the formal specification of the project Tokeneer), which is presented in Figure \ref{fig:schema}, requesting enrolment involves EnrolContext, KeyStore, AuditLog, Internal. Schemas in Z consist of an upper part, in which some variables are declared, and a lower part, which describes the relationship between values and variables. The notation $\Xi$ indicates an operation in which the state does not change, and the apostrophe indicates the state of the variable after the change\cite{Spivey89}. RequestEnrolment specifies that the ID station will request enrolment by displaying a request string on the screen and keeping the display blank. This is only possible while there is no Floppy present. Therefore, initially \eifkw{floppyPresence = absent} and \eifkw{enclaveStatus} set to \eifkw{notEnrolled}. An \eifkw{ensure} clause was used in the creation procedure to guarantee this after the initialization of \eifkw{ID\_STATION} object:

\begin{figure}
{\small
\[
\begin{array}{c@{\hspace*{15pt}}c}
\begin{array}{l}
\eifkw{make}\\
\hspace{0.5cm}\eif{\ldots Some lines were omitted\ldots}\\
\hspace{0.3cm}\eifkw{ensure}\\
\hspace{0.5cm}\eif{enclave\_status = cons\_floppy.not\_enrolled}\\
\hspace{0.5cm}\eif{floppy\_presence = cons\_internal.absent}\\
\hspace{0.5cm}\eif{token\_removal\_timeout = 0}\\
\hspace{0.3cm}\eifkw{end}
\end{array}
\end{array}
\]
} 
\caption{ensure clause in feature make}
\label{fig:ensure}
\end{figure}

Figure \ref{fig:invar} depicts the class invariant for class \eifkw{ID\_STATION}. It states that a message displayed on the display outside the enclave is one of the available from the list of messages (i.e. \eif{constants.display\_message.has(current\_display)}) and that class attribute \eif{constants} is attached to an object (i.e. \eif{constants /= Void}).

\begin{figure}
{\small
\[
\begin{array}{c@{\hspace*{15pt}}c}
\begin{array}{l}
\eifkw{invariant}\\
\hspace{0.5cm}\eif{constants.display\_message.has(current\_display)}\\
\hspace{0.5cm}\eif{constants /= Void}
\end{array}
\end{array}
\]
} 
\caption{Invariants of \eifkw{ID\_STATION} Eiffel class.}
\label{fig:invar}
\end{figure}


\begin{figure}
{\small
\[
\begin{array}{c@{\hspace*{15pt}}c}
\begin{array}{l}
\eifkw{feature}~ \eifcomment{Element Change}\\
\hspace{0.3cm}\eif{set\_current\_display(v: STRING)}\\
\hspace{0.5cm}\eifkw{require}\\
\hspace{0.7cm}\eif{constants.display\_message.has(v)}\\
\hspace{0.7cm}\eif{modify\_model("current\_display", Current)}\\
\hspace{0.5cm}\eifkw{do}\\
\hspace{0.7cm}\eif{current\_display := v}\\
\hspace{0.5cm}\eifkw{ensure}\\
\hspace{0.7cm}\eif{current\_display = v}\\
\hspace{0.5cm}\eifkw{end}
\end{array}
\end{array}
\]
} 
\caption{Feature equipped with \eifkw{modify} clause}
\label{fig:modify}
\end{figure}

Figure \ref{fig:modify} shows the implementation of procedure \eifkw{set\_current\_display}. Its first pre-condition was added to satisfy the invariant ensuring that argument \eif{v} belongs to the allowed displayed messages. The second pre-condition restricts the procedure to change values only to model query \eifkw{current\_display}.


Figure \ref{fig:idstation} shows the final version of class \eifkw{ID\_STATION}: with the respective annotations for Autoproof to successfully verify the class. In class \eifkw{ID\_STATION}, class attributes \eif{current\_screen} and \eif{current\_display} implements the physical screen and display, respectively, of the enclave.

\begin{figure}
{\small
\[
\begin{array}{c@{\hspace*{15pt}}c}
\begin{array}{l}
\eifkw{class}\\
\hspace{0.3cm}\eifkw{ID\_STATION}\\
\hspace{0.5cm}\eif{\ldots Some lines were omitted\ldots}\\
\\
\eifkw{create}\\
\hspace{0.3cm}\eif{make}\\
\\
\eifkw{feature}~ \eifcomment{Initialization}\\
\hspace{0.3cm}\eif{make}\\
\hspace{0.5cm}\eifkw{note}\\
\hspace{0.7cm}\eif{status: creator}\\
\hspace{0.5cm}\eifkw{do}\\
\hspace{0.7cm}\eifkw{create} \eif{ constants}\\
\hspace{0.7cm}\eif{current\_display := constants.blank}\\
\hspace{0.7cm}\eifkw{create} \eif{ current\_screen.make}\\\\
\hspace{0.7cm}\eifkw{create} \eif{ cons\_floppy}\\
\hspace{0.7cm}\eif{enclave\_status := cons\_floppy.not\_enrolled}\\
\hspace{0.7cm}\eif{token\_removal\_timeout := 0}\\\\
\hspace{0.7cm}\eifkw{create} \eif{ cons\_internal}\\
\hspace{0.7cm}\eif{floppy\_presence := cons\_internal.absent}\\
\hspace{0.5cm}\eifkw{ensure}\\
\hspace{0.7cm}\eif{enclave\_status = cons\_floppy.not\_enrolled}\\
\hspace{0.7cm}\eif{floppy\_presence = cons\_internal.absent}\\
\hspace{0.7cm}\eif{token\_removal\_timeout = 0}\\
\hspace{0.5cm}\eifkw{end}\\
\\
\eifkw{feature}~ \eifcomment{Element Change}\\
\hspace{0.3cm}\eif{set\_current\_display(v: }\eifkw{STRING})\\
\hspace{0.5cm}\eifkw{require}\\
\hspace{0.7cm}\eif{constants.display\_message.has(v)}\\
\hspace{0.7cm}\eif{modify\_model("current\_display", Current)}\\
\hspace{0.5cm}\eifkw{do}\\
\hspace{0.7cm}\eif{current\_display := v}\\
\hspace{0.5cm}\eifkw{ensure}\\
\hspace{0.7cm}\eif{current\_display = v}\\
\hspace{0.5cm}\eifkw{end}\\\\

\eifkw{feature}~ \eifcomment{Access}\\
\hspace{0.3cm}\eif{constants : }\eifkw{CONST}\\
\hspace{0.3cm}\eif{current\_screen : }\eifkw{SCREEN\_DISPLAY}\\
\hspace{0.3cm}\eif{current\_display : }\eifkw{STRING}\\
\eifkw{invariant}\\
\hspace{0.3cm}\eif{constants.display\_message.has(current\_display)}\\
\hspace{0.3cm}\eif{constants /= Void}\\
\eifkw{end}
\end{array}
\end{array}
\]
} 
\caption{Verified \eifkw{ID\_STATION} Eiffel class.}
\label{fig:idstation}
\end{figure} 

\section{Empirical Results}
\label{sec:empirical}

The usability of a verification tool cannot be considered in isolation and, in particular, cannot be hived off by the effectiveness of the tool itself. First, as a general observation, the cost of using an instrument can only be justified by its return, which can ultimately be linked to financial consideration by top management. Second, and this aspect is less general and more peculiar to the auto-active verification approach, a tool like AutoProof is as much effective and usable as is its ability to discharge verification conditions completely automatically, without feeding the code of annotation overhead or requiring particular tweaking. Finally, the necessity for users to add further annotations and dedicate extra effort (and considerable time) is, by itself, an obstacle to adoption and (technically) an usability issue. Verification tools should require minimal annotational effort and give valuable feedback when verification fails.

The case study analyzed in this paper presented good results in term of automatic discharge of verification conditions, though not comparable to others seen in literature \cite{Furia-PT15a}. In total there were 38 generated proof. Of these, 22 (58\%) were discharged automatically, 8 (21\%) could  not be satisfied, and the rest (21\%) failed due to internal errors, which in our case were basically caused by the attempt to create objects in the contract, and that is not allowed by the tool. As observed before, the success of verification is unsurprisingly linked to the complexity of programs \cite{Furia-PT15a}. Previous literature mostly dealt with students users and university projects. The use of Tokeener as a benchmark demands for detailed comparisons with different verification efforts (for example, \cite{ rivera:tokeneer:16}).

\section{Related Work}
\label{sec:related}
Formal/mathematical notations have existed for a long time and have been used to specify and verify systems. Examples are process algebras \cite{Baeten05}, specification languages like Z \cite{Abrial80}, B \cite{Abrial05} and Event-B \cite{Abrial10}. The Vienna Development Method (VDM) is one of the earliest attempts to establish a formal method for the development of computer systems \cite{Jones80}. A survey of these (and others) formalisms can be found in \cite{Mazzara10} while a discussion on the methodological issues of a number of formal methods is presented in \cite{Mazzara09}.

All these approaches (and others described in the literature) still leave an open issue, i.e., they are built around strict formal notations which affect the development process from the very beginning. These approaches demonstrate a low level of flexibility. To overcome this problem, a seamless methodological connection built on top of a portfolio of diverse notations and methods is presented in \cite{Gmehlich13}. Another approach is presented in \cite{rivera:tokeneer:16,rivera:eb2java:15} using \cite{rivera:eb2java:14}, where users start the development of system from a strict formal notation (i.e. Event-B), to then automatically translate it to Java code with JML \cite{Leavens:jml:2006} specifications embedded (following Design-by-Contract methodology). Even though this approach enables users with less mathematical expertise to work on formal development, it does not give a seamlessly methodology for the development as presented in this paper.

On the other side, Design-by-contract\cite{Meyer:OOSC:97} when combined with AutoProof technology offers the pros of both rigorous methodologies and supporting tools able to semi-automate the process. Before this to be available for the average developer it is however necessary to improve the users' experience. A comparison between different approaches (for example Event-b/Rodin and Design-by-contract/AutoProof) is beyond the scope of this paper and it is left as future work. 

\section{Conclusion}
\label{sec:conclusion}
AutoProof allows for ``static debugging'', i.e. debugging becomes possible without the need of executing the program. The most effective way to release correct software is a combination of static debugging and traditional run-time debugging. Being all human activities (therefore including programming and testing itself) error-prone, there is no magic or free lunches out there. Abandoning testing and adopting a proof-oriented approach does not make miracles, debugging remains a trial-and-error long and laborious process. AutoProof does not change the rules of the game: developers will have to try, observe the results and make changes as a consequence. A proof-oriented approach does not make the process smoother and necessarily simpler. However, it makes it more accurate and robust, therefore effective. Adjustment can be now focused on the the implementation side (possibly sinergically with run-time debugging), on the specification side (the contracts used to annotate the code as integral part of the code itself), or in the proof itself (fine-tuning may be necessary for AutoProof and its behind-the-curtains machinery to be able to prove correctly).

All this comes with a cost: the willingness and ability of the user to use extra tools and being able to master them, and possibly invest extra time in the process. On the other side, it is necessary for the tools to be simple to master and to provide intelligible feedback. 


The Tokeener project case study showed the efficacy of AutoProof in verifying a real piece of software, the complexity of which can be compared not only with most of the commercial Off-the-Shelf software, but also with safety and financial-critical applications, both in terms of computational logic and architectural organization. AutoProof is capable to verify industrial software and may well be adopted in commercial companies and its use injected into the development process. However, some obstacles have been identified that could prevent its broader adoption. 
 
As result of an academic effort, documentation is not at par with commercial software, in particular for what concerns the size of the library of correctly verified examples: tutorials on the official website are quite useful, but not enough. On top of this, the tool itself has limitations. First, existing implementations need to be modified in order to be verified. This would represent an unsurmountable obstacle in most institutions since the overall cost of code adaptation may overrun the saves occurring to the testing phase. This consideration may be different, however, for safety-critical systems. Second, the Eiffel IDE - necessary for functioning - calls for increased stability and robustness.

\section*{Acknowledgments}
We would like to thank Innopolis University for logistic and financial support, and the laboratories of Software Engineering (SE) and Service Science and Engineering (SSE) 
for the intellectual engagement and vivid discussions.

\bibliographystyle{ieeetr}


\end{document}